\documentclass[12pt,epsf]{article}
\usepackage{sc3conf}
\usepackage{epsfig}
\def\be{\begin{equation}}
\def\ee{\end{equation}}
\def\bea{\begin{eqnarray}}
\def\eea{\end{eqnarray}}
\def\ba*{\begin{eqnarray*}}
\def\ea*{\end{eqnarray*}}
\def \ee {\epsilon}

\def \frac#1#2{{ #1 \over #2}}

\def \ee {{\rm e}}

\def \ee {\epsilon}

\def\be{\begin{equation}}
\def\ee{\end{equation}}

\begin{document}
\raggedbottom

\title{Dark Energy May Probe String Theory}

\authors{L.Mersini and M. Bastero-Gil\adref{1,2,3}
  }

\addresses{\1ad  Scuola Normale Superiore, Piazza dei Cavalieri, Pisa,
  Italy, 
\nextaddress \2ad Physics Department, Syracuse University, Syracuse,
  NY 13244, USA.\\ SU-GP-02/11-3, SU-4252-772 
\nextaddress \3ad Centre for theoretical physics, University of
  Sussex, Brighton, BN1 9QJ, UK
   }

\maketitle


\begin{abstract}
   The problem of dark energy arises due to its self-gravitating
properties. Therefore explaining vacuum energy may become a question 
for the realm of quantum gravity,
that can be addressed within string theory context. In this talk I
concentrate on a recent, string-inspired model, that relies on
nonlinear physics of short-distance perturbation modes, for explaining
dark energy without any fine-tuning. 
Dark energy can be observationally probed by its equation of
state, w. Different models predict different types of equations of
state and string-inspired ones have a time dependent w(z) as
their unique signature. Exploring the link between dark energy and
string theory may provide indirect evidence for the latter, by means
of precision cosmology data.  
\end{abstract}

\section{Introduction}

There is still no fundamental physical theory of the very early
universe which addresses issues that arise from the regime of
transplanckian physics. It is the lack of a fundamental theory, valid
at all energies,  that 
makes the model building of the transplanckian regime very
interesting. The main issue is how much are the known
observables affected by the unknown theory. 
The apparently {\it ad hoc} modification of the dispersion relation
at high energies is contrained by the criterion that its low energy
predictions do no conflict the observables.

In this
talk we address the possibility that transplanckian regime
can contribute to the observed dark energy of the universe through a
counterintuitive UV/IR mixing\cite{mbk}. The physics mechanism that
gives rise to dark energy is the freeze-out of ultralow frequency
short distance modes by the expansion of the background universe. This
transplanckian dark energy (TDE) model is motivated from closed string
theory in a toroidal geometry by invoking superstring duality. 

The transition from string theory to conventional cosmology is
becoming increasingly important to theoretical physics. As we describe
in \cite{bfm} corrections to short distance physics due to the
nonlocal nature of strings contribute to dark energy. The possibility
to detect their signature observationally is thus very intriguing.  

A potential way for detecting dark energy and distinguishing it from a
pure vacuum energy $\Lambda$ is through its equation of state. I will
briefly touch upon the calculation of the stress-energy tensor, in
order to address the puzzle of ``cosmic coincidence'' (see \cite{bm} for
details), and the time dependence in the equation of state $w(z)$
predicted by this model. 

\section {The TDE Model}

Let us start with the generalized Friedmann-Lemaitre-Robertson-Walker
(FLRW) line-element in the presence of scalar and tensor
perturbations. For simplicity, we will take the class of inflationary
scenarios that has a power law solution for the scale factor
$a(\eta)$, $a(\eta)=|\eta_c/\eta|^{\beta}$, where $\beta \geq 1$ and
$|\eta_c|= 
\beta/H(\eta_c)$. The initial
power spectrum of the perturbations can be computed once we solve the
time-dependent equations in the scalar and tensor sector. The mode
equations for both sectors reduce \cite{mode1, tp} to a
Klein-Gordon equation of the form
\begin{equation}  
\mu_n^{\prime \prime} + \left[ n^2 - \frac{a^{\prime \prime}}{a} \right]
\mu_n=0 \,,
\label{kg}
\end{equation}
where the prime denotes derivative with respect to conformal time,
and $n$ is the comoving wavenumber.
Therefore, studying perturbations in a FLRW background is equivalent
to solving the mode equations for a scalar field $\mu$ related
(through Bardeen variables) to the perturbation field  in the expanding
background. The above equation represents a linear dispersion relation
for the frequency $\omega$,
\begin{equation}
\omega^2= p^2 = \frac{n^2}{a^2} \,.
\label{ldis}
\end{equation}
The dispersion relation of Eq. (\ref{ldis}) holds for values of
momentum smaller than the Planck scale $M_P$. There is no reason to believe
that it remains linear at ultra-high energies larger than $M_P$.

In what follows, we replace the linear relation $\omega^2(p) = p^2
=n^2/a(\eta)^{2}$ with a nonlinear dispersion relation
$\omega(p) =F(p)$. Therefore, in Eq.  (\ref{kg}), $n^2$ should be
replaced by: 
\begin{equation}
n_{eff}^2 = a(\eta)^2 F(p)^2 = a(\eta)^2 F[n/a(\eta)]^2 \,.
\label{neff}
\end{equation}

Let us consider the following
Epstein function \cite{epstein} for the dispersion relation, in the
squared bracket in Eq. (\ref{kg}):
\begin{eqnarray}
\omega^2(p) -\frac{a^{\prime \prime}}{a}&=& F^2(p)= p^2
\left(\frac{\epsilon_1}{1+ e^x} + 
\frac{\epsilon_2 e^x}{1+e^x} + \frac{\epsilon_3 e^x}{(1+e^x)^2}\right)
\,,\label{omega}\\ 
n^2_{eff} -\frac{a^{\prime \prime}}{a}&=& a^2(\eta) F^2(n,\eta)= n^2
\left(\frac{\epsilon_1}{1+ e^x} + 
\frac{\epsilon_2 e^x}{1+e^x} + \frac{\epsilon_3 e^x}{(1+e^x)^2}\right) \,,
\label{disp}
\end{eqnarray}
where $x=(p/p_C)^{1/\beta}=  A |\eta|$, with $
A=(1/|\eta_c|)(n/p_C)^{1/\beta}$, in general for power-law inflation
with $\beta \geq 1$. This is the most general expression
for this family of functions.  For our purposes, we will constrain
some of the parameters of the Epstein family in order to satisfy
the features required for the dispersion relation  as
follows.  

Imposing the requirement of superstring duality fixes $\epsilon_2$, and
the condition of a nearly linear dispersion relation for 
$p<p_C$ requires that
\bea
\epsilon_2=0 \,,\;\;\;\;\;\;
\frac{\epsilon_1}{2} + \frac{\epsilon_3}{4}=1 \,.
\eea
\begin{figure}[t]
\epsfxsize=10cm
\epsfxsize=10cm
\hfil \epsfbox{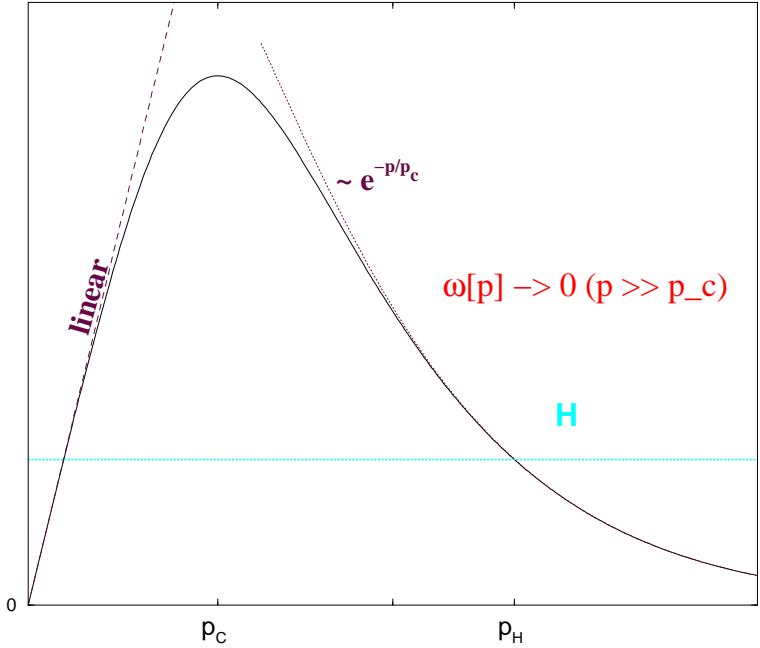} \hfil
\caption{{\footnotesize Shown is our family of dispersion relations,
for $\beta=1$ and representatives values of $\epsilon_1$ (solid
lines). We have also included the linear one
(dotted line) for comparison.} }
\label{fig1}
\end{figure}
Still we will have a whole family of functions parametrised by the
constant $\epsilon_1$, as can be seen in Fig. \ref{fig1}.

The initial vacuum state is well-defined in this model and it is a
Bunch-Davis vacuum, due to the dispersion function going
asymptotically flat at early times. At late times the 
solution becomes a squeezed state by mixing of positive and negative
frequencies:
\begin{equation}
\mu_n \rightarrow_{\eta \rightarrow +\infty} \frac{\alpha_n}{\sqrt{2
\Omega^{out}_n}} e^{-i\Omega^{out}_n \eta} + \frac{\beta_n}{\sqrt{2
\Omega^{out}_n}} e^{+i \Omega^{out}_n \eta} \,,
\end{equation}
with $|\beta_n|^2$ being the Bogoliubov coefficient equal to the
particle creation number per mode $n$, and $\Omega^{out}_n \simeq
\sqrt{\epsilon_1} n$. Using the linear transformation
properties of hypergeometric functions we
obtain:
\begin{equation}
|\beta_n|^2 = \frac{e^{-2\pi \sqrt{\epsilon_1}}}{2 \sinh 2\pi
 \sqrt{\epsilon_1}} 
\label{betak2}\,.
\end{equation}
Clearly the spectrum of created particles is nearly thermal to high
accuracy. 

We define the $tail$ as the range of those transplanckian modes in
fig.1  whose frequency is less or at  most
equal to the present Hubble rate, $H_0$ (see Fig. 2). Since $H$
has been a decreasing function of time, many modes, even those in the
ultralow frequency range, have become dynamic and redshifted away one
by one, everytime the above condition is broken, i.e. when the
expansion rate $H$ dropped below their frequency. Clearly, what is
left from the tail modes at the present Hubble scale have always been
frozen. They contain vacuum energy of very short distance, hence of very low
energy. The last mode in the tail (at infinite momenta) would decay
when and only if $H=0$ by acquiring a kinetic term. The $tail$ starts
from some value $p_H$ which must be found by solving the equation
\begin{equation}
\omega^2(p_H)=H^2_0 \,.
\label{omegah}
\end{equation}
The range of the modes defining the  $tail$ is then for $p_H<p<\infty$.
Their equation of state depends on the
evolution of $H$ and is a complicated tracking solution because they
contribute to the expansion rate for $H$ and their equations of motion
are coupled to the Friedmann equation for expansion.

\begin{figure}[t]
\epsfxsize=10cm
\epsfxsize=10cm
\hfil \epsfbox{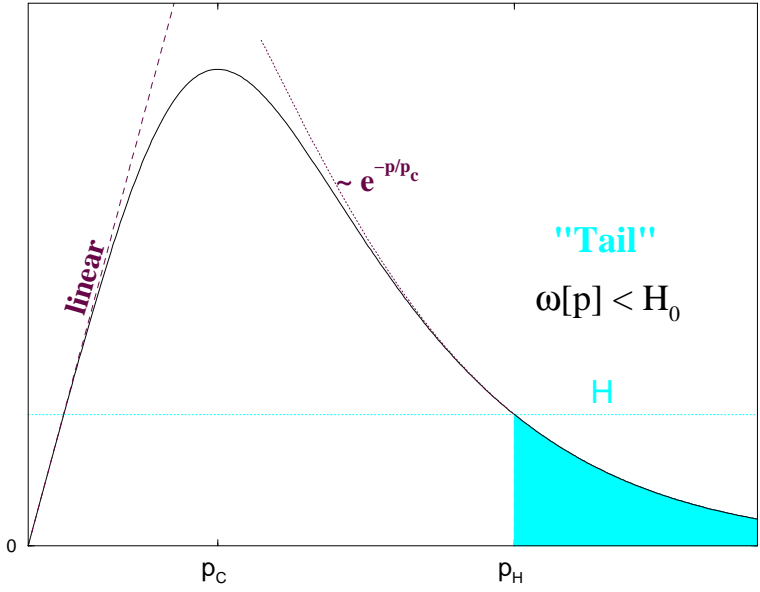} \hfil
\caption{{\footnotesize The range of modes in the tail, $p_H < p <
\infty$, defined by Eq. (\ref{omegah}). $H_0$ is the present value of
the Hubble constant.} }
\label{fig2}
\end{figure}

However let us first calculate their energy contribution today,
by asking which ``tail'' modes have always been frozen from
inflationary times to present and how much energy do they contain. 
The energy for the tail is given by:
\begin{equation}
\langle \rho_{tail} \rangle = \frac{|\beta_n|^2}{2 \pi^2 } \int^{\infty}_{p_H}
\frac{\omega(p)}{p} \omega(p) p^2 dp   \label{entail}\,,
\end{equation}
while the expression for the total energy is:
\begin{equation}
\langle \rho_{total} \rangle = \frac{|\beta_n|^2}{2 \pi^2 }
\int^{\infty}_{0} \frac{\omega(p)}{p} \omega(p) p^2 dp \,.
\end{equation}
The numerical calculation of the tail energy produced the following
result: for random  different values of 
the free parameters, the dark energy of the tail is
$\rho_{tail}=10^{-122} f(\epsilon_1)$, times less than the total energy
{\em during inflation},
i.e. $\frac{\rho_{tail}}{\rho_{total}}=10^{-122} f(\epsilon_1)$ at 
Planck time. The prefactor $f(\epsilon_1)$, which depends weakly on the
parameter of the dispersion family $\epsilon_1$, is a small number
between 1 to 9, which clearly can contribute at the most by 1 order of
magnitude. 
{\em This is an amazing result!} The lack of {\em fine-tuning} in
obtaining the right ammount of dark energy can be understood from the
following feature: due to the decaying exponential, the main
contribution to the energy integral in Eq.
(\ref{entail}) comes from the highest value of this exponentially
decaying frequency, which is the value of the integrand at the tail
starting point, $p_H \sim O(M_P)$, i.e., 
\begin{equation}
\langle \frac{\rho_{tail}}{\rho_{total}} \rangle \approx
\frac{p_H^2}{M_P^4} \omega^2(p_H) \approx \frac{H_0^2}{M_P^2}\approx
10^{-122} \,.  
\end{equation} 
From the physical requirement that the tail modes must have always been
frozen, the tail starting frequency $\omega(p_H)$ is then proportional to
the current value of Hubble rate $H_0$ (Eq. (\ref{omegah})).

\section{TDE Dispersion Relation from Closed Strings in Toroidal Cosmology}

Here we attempt to motivate the exponentially supressed transplanckian
dispersion function from T-duality of the following string theory
model. 
Let us consider the Brandenberger-Vafa model \cite{vafa} of a 
D-dimensional anisotropic torus with radius $\bar{R_i}$, by including
the dynamics of both modes: momentum modes, $p_{1,i}=m/\bar R_i$
(where $m$ is the 
wavenumber), and winding modes with momenta $p_{2,i}=w\bar
R_i/\alpha^\prime$. The 
dimensionless quantity for the radius is
$R_i=\bar{R_i}/\sqrt{\alpha^\prime}$, where $\alpha^\prime$ is 
the string scale. Based on the arguments presented in \cite{vafa}, we
{\it choose} a 
cosmology with three toroidal radii equal and large $R \gg 1$ in
units of the string or Planckian scale, with 
the other $(D - 3)$
toroidal radii equal and small $R_C \ll 1$. Here the subscript $C$
refers to compactified dimensions.  
Then,  $R(t)$ becomes the scale factor
for the 3+1 metric in conventional FRW 
cosmology $R(t)=a(t)$, while $R_C$ corresponds to
the radius, in this factorizable metric, of the $D-3$ compact
dimensions $z_j$ that decrease with time,
\begin{equation}
d s_D^2 = -dt^2 + 4 \pi R^2(t) dx_i^2 + 4 \pi R^2_{C}(t) dz_j^2 =
a(\eta)^2 [- d\eta^2 + dy^2] + d s_{D-3}^2\,. 
\label{metric2}
\end{equation}
These solutions were found by Mueller, Ref. \cite{mueller}, by
ignoring the backreaction of string matter on the geometry. 

The partition function for this system was  calculated, from first
principles, by summing up over their momenta in \cite{kikkawa}:
\begin{equation}
Z= \sum_\sigma e^{-n_\sigma \epsilon_\sigma}\,,
\end{equation} 
where $n_{\sigma}$ is the number of strings in 
state $\sigma$ with energy $\epsilon_{\sigma}$
\begin{equation}
\epsilon_\sigma=p_0=\sqrt{\left(\frac{m}{R}\right)^2 + (w R)^2 + N +
\tilde{N} -2 } \,,
\label{epsilon}
\end{equation}
and $\sigma$ counts over $(m,\,w)$, with the constraint $N-\tilde{N}=m
w$ for closed strings where $N$ and $\tilde{N}$ are the
sums over the left- and right- mover string excitations, respectively.
By now, in Eq.(\ref{epsilon}),
we are considering only the large 3 spatial dimensions.
The string state can also be described by 
its left and right momenta, $k_L = p_1+p_2$, $k_R=p_1 - p_2$. The string
state for left and right modes can be  
expanded in terms of the creation and annihilation operators
$\alpha_m$, $\tilde\alpha_n$, with higher excitation string states
given by $N=\sum_{n=1}^{\infty} \alpha_{-n}\alpha_n$ (similarly
for $\tilde{N}$),  and string energy 
$L_0 + \tilde{L}_0 = p_1^2 +p_2^2 + (N +\tilde{N} -2)/\alpha^\prime$.   

We would like to write the path integral for this configuration in
terms of quantum fields\footnote{Below we use quantum string equations
under the assumption that the dilaton is massive and stable.}. The
path integral is calculated from the hamiltonian density.   
The hamiltonian density over the
fields in configuration space is extracted from the string spectrum in
such a way that its Fourier transform 
in $k$-space corresponds to the string energy expression Eq. (\ref{epsilon}). 

Our system of winding and momentum modes is described by nonequilibrium
dynamics due to the expanding background spacetime. All the
information about the evolution of these modes will be contained in
the effective action. The kinetic terms are unambiguous while for the
interaction terms we must appeal to simplicity and the requirement of
T-duality.  

Correlation functions are obtained by using the correspondence between
the Euclidean path integral of the persistence vacuum amplitude
$|\langle in | out \rangle|^2$ and the partition function $Z$.  
All the string quantum operators below are promoted to quantum field
operators with the corresponding hamiltonian density ${\cal H}(t,x)$ in
configuration space derived from the quantum string hamiltonian
${\cal H}(t)$. 

The following calculations are done in the conformally flat 
background  through the scaling of the 
fields and operators with the conformal factor $a(\eta)$.
The momentum field $\phi_1(R,x)$ and the winding field
$\phi_2(R,x)$ are defined by the relation:
\begin{equation}
\phi_i(x)= \int e^{i p_i x} \phi_i(p_i) d^3p_i\,, \;\;\;\int |\nabla
\phi_i|^2 d^3x= \int d^3p_i p_i^2 \phi_i(p_i) \,,  
\end{equation}  
where
\begin{equation}
\nabla= R \partial/\partial x \ = \partial/\partial y \,,
\end{equation} 
and $p_i=p_1,\,p_2$. 
Let us also define two new fields, $\psi_L(R,x)$ and $\psi_R(R,x)$,
with momenta $k_L,\,k_R$ that are the left and right combinations of the
Kaluza Klein momentum and winding modes 
\begin{eqnarray}
\psi_L(R,x) &=& \phi_1(R,x) + \phi_2(R,x) \,, \label{psi1}\\
\psi_R(R,x) &=& \phi_1(R,x) - \phi_2(R,x) \,. \label{psi2}
\end{eqnarray}
These fields live in the expanding (3+1) spacetime
dimensions.

The Hamiltonian density ansatz that would describe the energy of our 
two string states in the $D=3$ expanding dimensions 
with energy $H = L_0 +\widetilde L_0$, 
including the  oscillators from string's higher excitations $(N
+\widetilde N -2)/\alpha^\prime$,    
is similar to the hamiltonian of spin waves in a periodic
lattice\footnote{Torus is obtained by identifying the first and the  
last lattice sites, thus the periodicity.}. Our lattice spacing is
given by the string scale $\sqrt{\alpha'}$. Therefore the hamiltonian
density can be written for this dual lattice in terms of wave
functional ``spin'' fields $\psi_L(R,x),\psi_R(R,x)$ 
of Eqs. (\ref{psi1}), (\ref{psi2}) as follows

\begin{equation} 
{\cal H}_3 = |\nabla \psi_L|^2 + |\nabla \psi_R|^2 |+ |\nabla
\psi_L||\nabla \psi_R|+ m_0^2 ( |\psi_L|^2 + 
|\psi_R|^2) + g_1 (|\psi_L|^4 + |\psi_R|^4) + g_2 |\psi_L|^2
|\psi_R|^2 \,,
\end{equation}
where the fields $\psi_L,\,\psi_R$ are expanded in terms of the mode
functions $u_n,\tilde u_n$, 
\begin{equation}
\psi_L = \Sigma u_n b_n + u_n^* b_n^+\,, \;\;\;\; \psi_R = \Sigma
\tilde u_n \tilde b_n + \tilde u_n^* \tilde b_n^+\,,
\end{equation}
and  $b_n,\tilde b_n$ are the normalised quantum creation and
annihilation operators of $\alpha_n, \tilde \alpha_n$. 
The commutation relation for the unnormalised operators are such that
$[\alpha_n, \alpha_m^+] = \omega_{\pm}\delta_{nm}$ with 
$\omega_{\pm}$ the frequency of left and right moving modes. 

The periodic lattice condition $N-\widetilde N=mw$ introduces  
an interaction term in the hamiltonian ${\cal H}_3$ of 
the form $\nabla\psi_L\nabla \psi_R$. 
In terms of the 2-component state $\Psi_{EN}
= (\psi_L, \psi_R)$, the hamiltonian  reads,
\begin{equation}
{\cal H}_3= |\nabla \Psi_{EN}|^2 + \nabla \Psi_{EN} \widehat{X} \nabla
\Psi_{EN} +  
m_0^2 |\Psi_{EN}|^2 +
g_1 |\Psi_{EN}|^4  + (g_2 - 2 g_1) |\Psi_{EN} \widehat{X}
\Psi_{EN}|^2 \,, 
\end{equation}
with 
\begin{equation} 
\widehat{X}= \left( \begin{array}{cc} 0 & 1/2 \\ 1/2 & 0 \end{array}
\right) 
\end{equation}
The system is known as the dual momentum-space lattice, and for $g_2=2
g_1$ reduces 
to the XYZ model of condensed matter. Let us for simplicity limit to
the XYZ model case, $g_2=2 g_1$, for the rest of this talk. 

These periodic lattice systems studied in 3+1 dimensions in terms of
Bloch wavefunctions have a solution which respect lattice translation 
invariance, $exp(-p l)$, with the lattice spacing ``$l$''  
equal to the string scale $\sqrt{\alpha^\prime}$. 
The interaction term,
in the tight-binding approximation, lifts the degeneracy between  
the energy eigenstates due to the leakage/tunnelling of the
wavefunction from one lattice site to the neighbour site. 
As a result the gap energy produced between
the ground (bound) state and higher excitation states is 
\begin{equation}
p^2 \Delta_p = p^2 |\cos(2 \theta)|= p^2 ~ |2 cos^2 (\theta) - 1| \,,
\label{gap}
\end{equation}
in which 
\begin{equation}
p l =p \sqrt{\alpha^\prime}= \sqrt{\alpha^\prime(p_1^2 + p_2^2)}=
\sqrt{(\frac{m}{R})^2 + (wR)^2} \,,
\end{equation} 
and $\theta \rightarrow \theta + i p l$. Therefore, 
\begin{equation} 
\Delta_p \leq 2 \cosh^2(p l) -1 \,. 
\label{massgap}
\end{equation}
The first term in $\theta$
is a pure phase of rotation of the ``spin-wave'' in the dual lattice,
but the second term describes the tunnelling of the wavefunction to the
nearest neighbour\footnote{ In condensed matter this is known as
Coulomb dipole type of vortex interaction.}.
The gap energy of Eq. (\ref{gap}) introduces a correction to the
kinetic energy, such that in momentum space the hamiltonian reads
\begin{equation} 
{\cal H}_3= z_p p^2 |\Psi_{EN}|^2 + m_0^2 |\Psi_{EN}|^2 + g_1
|\Psi_{EN}|^4 \,,
\end{equation} 
with $z_p = 1 + \Delta_p$. 
One can evaluate the reduced 3 dimensional partition
function $Z$ from the above hamiltonian\cite{bfm}. We want to 
separate our modes into system (S) + environment (E) degrees of freedom, and
coarse grain by integrating out the degrees of freedom for the
environment. This amounts to finding out the backreaction of the
coarse grained environment on the system, and
eventually leads to the RGE's\footnote{The running of the coupling
constants with time depends on how one selects the
environment and the system.}. We will consider as
environment all the short wavelength 
modes with momenta 
\begin{equation}
(E): \;\;\;\;\; \frac{\Lambda}{b} < p^E
=\frac{1}{\sqrt{\alpha^\prime}}[(m/R)^2 + (w R)^2]^{1/2} < \Lambda \,, 
\label{env}
\end{equation} 
where the cutoff $\Lambda= (\alpha^\prime)^{-1/2}$ is the string
scale because $(\alpha^\prime)^{1/2}$ is identified with the lattice spacing
$l$, and $b=a(t)/a(t_0)$ is the coarse grain scaling parameter, where 
$t_0$ is the initial time. The scale factor $a(t)$ plays the 
role of the collective coordinate describing the environmental degrees
of freedom. Time in this procedure is playing the role of a scaling
parameter and dynamics is being replaced by scaling, an
artificial procedure, known as Kadanoff-Migdal transform, that relates 
the microscopic and macroscopic properties of a system based on the
existence of scaling properties of the system in the infrared
limit. 

The system modes are the ones with:
\begin{equation}
(S): \;\;\;\;\; p^S< \frac{\Lambda}{b} \,, 
\label{system} 
\end{equation}
>From the above definitions of system and environment,
Eqs. (\ref{system}) and  (\ref{env}), 
at initial times when $b\approx 1$ we have $p \le \Lambda/b$ thus all
our modes, momentum and winding, are in the system;   
but at later times when $b \geq 1$ more and more winding modes
systematically transfer  
to the environment because the condition of
Eq. (\ref{system}), $p=\frac{1}{\sqrt{\alpha^\prime}}[(m/R)^2 + (w
R)^2]^{1/2} \le 
\Lambda/b$ is satisfied only for vanishingly small winding numbers
$w \rightarrow 0$. As $t$ becomes large, the system contains
$m \leq R \Lambda$, $w=0$, i.e.  
all the modes except $m \leq R \Lambda, w =0$ have been transfered to
the environment.

After integrating out the high energy environmental modes in the above
action, one can extrapolate the correlation function from the coarse
grained effective action of the system. 
The canonical two-point correlation function at high energy for 
system-environment interaction is calculated from the path 
integral of the canonical fields $\tilde{\Psi}_{S,E}$ in
momentum space (Fourier transform of $G^{\Lambda/b}$). 
It is related to the correlation function of the original fields $\Psi_E$ 
(which decreases at high energy) as follows 
\begin{equation}
\langle \Psi_E \Psi_E \rangle= 
\frac{\langle \tilde \Psi_E \tilde \Psi_E \rangle}{z_p} 
\,. 
\label{corrfunct}
\end{equation}
where $z_p = 1 + \Delta_p$ and $\Delta_p$ is given in Eq.(\ref{massgap}).
This is the crucial result for the 
interpretation of the cosmological dark energy.
Because of the mass gap, the correlation function is suppressed exponentially
in p-space. It is very familiar that a mass gap leads to an exponential
fall off in x-space, but here for the dual lattice the exponential fall off
is in momentum space. This may be traced to the T-duality of the closed 
strings and the resultant interchange of the IR/UV limits.
Thus the two point function of the original fields (by dividing with
the $z_p$ normalisations factor), goes as an exponential for large
momentum $p$, and as a polynomial for low momenta.

The observed small value $\Lambda \sim 10^{-120}$ in natural units
has an explanation in the toroidal cosmology of closed strings and thus the
dark energy provides an exciting opportunity to connect string theory to
precision cosmology.
We may argue that numerically the
size of the cosmological constant in the present approach
is a combination of the string scale
and the Hubble expansion rate in the sense that
$\Lambda/M_{Planck}^4 \simeq 10^{-120} \simeq (H_0/M_{Planck})^2$.
Therefore the correct amount
of dark energy obtained by this frequency dispersion
function does not require any fine tuning and relies,
besides a physical mechanism (such as freeze-out),
only on the string scale as the parameter of the theory.

\section{TDE Equation of State}

Nonlinearity of short distance physics and the breaking of Lorentz invariance results in a violation of Bianchi identity for transplanckian models. Therefore one
needs to modify the stress-energy tensor ($T_{\mu\nu}$) in the Einstein equations, such that the
modified ones satisfy Bianchi identity.

Based on the equation of motion as our sole information for
short-distance physics, we therefore use a kinetic theory approach for
modifying Einstein equations in the absence of an effective lagrangian
description. The assumption made is that a kinetic theory description
of the cosmological fluid is valid even in the transplanckian
regime. Despite its nonlinear behavior at short distances, this
imperfect fluid shares the same symmetries, namely homogeneity and
isotropy, as the background FRW universe. Then the corrections
$\tau_{\mu\nu}$ to the stress energy 
tensor $T_{\mu\nu}$ will also be of a diagonal form 
\be
\tau_{\mu\nu} = (\bar{\epsilon} + \Pi) u_{\mu}u_{\nu} + \Pi g_{\mu\nu}
\ee
In a similar manner to particle creation cases in imperfect
fluids \cite{zimdahl}, the highly nontrivial time-dependence of the
mode $p_H$ and the transfer of energy between regions, due to the
defrosting of this mode across the boundary $p_H$, gives rise to
pressure corrections in the fluid energy conservation law. The
defrosting of the modes results in a time-dependent ``particle number''
for regions near $p_H$. From kinetic theory we know that this
``particle creation'', (the defrosting of the modes), gives rise to {\em
effective viscous pressure modifications} \cite{zimdahl}. The
term $\Pi$ denotes the effective viscous pressure modification to the ``bare''
pressure, $\langle \bar p\rangle$.  

The criteria we will use for modifying $T_{\mu\nu}$ is that Bianchi
identity must be satisfied \cite{lorentzinvar} with the new expressions for
pressure, $P$,
\be \Sigma [ \dot \rho + 3 H ( \rho + \bar{p} + \Pi )] =
\Sigma_i [ \dot \rho + 3 H ( \rho + P)] = 0 \,,  
\ee
The presence of pressure corrections, $\Pi_i$, (where $i=II,H$ denotes 
transplanckian defrosted and tail modes respectively) is due to the 
exchange of energy between the two regions, from the defrosting of 
the modes $p_H$ at the boundary. This is directly related to the time 
dependence of the boundary $p_H$, which in turn is going to be 
controlled by the Hubble parameter $H$. In essence, there is an 
exchange of modes between the two regions. Although the number of
particles\footnote{We are loosely using the term  
particle here to refer to the wavepackets of the transplanckian 
modes, centered around a momenta $p_i$.} in each of these regions, 
$N_{II}$ and $N_H$, is not conserved, their rate of change, in the 
physical FRW Universe, is  
related through the conservation of the total number of particles 
which contains both of these components   
\be
\dot N_T = 0\,, \label{dotNT}
\ee

The contribution terms to pressure, $\Pi_i$, are related to decay
rates $\Gamma_i$ through \cite{zimdahl} 
 \be 
3 H \Pi_H = - ( \rho_H + \bar{p}_H)\Gamma_H \,,
\label{PiH}
\ee

Details of the calculation can be found in\cite{bm}. The value of the tail's decay rate $\Gamma_H$ is: 
\be \Gamma_H \simeq 3H + \frac{\dot H}{H} \simeq 3 H
(\frac{1 - w_{total}}{2})\,, \label{gammah} 
\ee where 
$w_{total}=\bar{p}_{total}/\rho_{total}$. Notice that $\Gamma_H$ is
positive for 
all equations of state $w_{total} \le 1$ and thus it slows down the
dilution of the tail with the scale factor.

The effective equation of state resulting from the modification
$\Gamma_H$ of stress-energy tensor becomes: 
\be 
w_H = (\frac{w_{total}-1 }{2})\,.
\ee
This is a tracking solution. During a radiation dominated universe,
$w_H = -1/3$ and only after matter domination, the tail starts
acquiring a strongly negative equation of state until it
asymptotically aproaches $w_H =-1$ when it becomes the only component
left in the expansion equation. 
Therefore cosmic coincidence is explained naturally by the intrinsic
time evolution of the tail which tracks the background equation of
state $w_{total}$ through its coupling to the Hubble parameter. 

Careful analysis of the highly anticipated new data from CMB, LSS and
SN1a will reveal more scrutinity between various
models\cite{alessandro}. 

\section{Acknowledgment}
LM issupported in part by the National
Science Foundation (NSF) under grant PHY-009412
and by the U.S. Depart
ment of Energy under contract number DE-FG02-85
ER40231.


\end{document}